\documentclass{iopart}
\usepackage{graphicx}
\usepackage{subfigure}

\jl{6}        
\eqnobysec    

\def\beq{\begin{equation}}
\def\eeq{\end{equation}}
\def\rmd{{\rm d}}
\def\bra{\big<}
\def\ket{\big>}

\begin{document}

\title[Massless Dirac particles in the vacuum C-metric]
{Massless Dirac particles in the vacuum C-metric}

\author{
Donato Bini$^*$,
Eduardo Bittencourt${}^\S {}^\ddag$ and
Andrea Geralico$^*$ 
}
\address{
  ${}^*$\
Istituto per le Applicazioni del Calcolo ``M. Picone'', CNR I-00161 Rome, Italy
}
\address{
  ${}^\S$\
CAPES Foundation, Ministry of Education of Brazil, Bras\'ilia, Brazil
}
\address{
  ${}^\ddag$\
Physics Department, \lq\lq La Sapienza" University of Rome, I-00185 Rome, Italy
}

\begin{abstract}
We study the behavior of massless Dirac particles in the vacuum C-metric spacetime, representing the nonlinear superposition of the Schwarzschild black hole solution and the Rindler flat spacetime associated with uniformly accelerated observers. Under certain conditions, the C-metric can be considered as a unique laboratory to test the coupling between intrinsic properties of particles and fields with the background acceleration in the full (exact) strong-field regime. The Dirac equation is separable by using, e.g., a spherical-like coordinate system, reducing the problem to one-dimensional radial and angular parts. Both radial and angular equations can be solved exactly in terms of general Heun functions. We also provide perturbative solutions to first-order in a suitably defined acceleration parameter, and compute the acceleration-induced corrections to the particle absorption rate as well as to the angle-averaged cross section of the associated scattering problem in the low-frequency limit. Furthermore, we show that the angular eigenvalue problem can be put in one-to-one correspondence with the analogous problem for a Kerr spacetime, by identifying a map between these \lq\lq acceleration'' harmonics and Kerr spheroidal harmonics.
Finally, in this respect we discuss the nature of the coupling between intrinsic spin and spacetime acceleration in comparison with the well known Kerr spin-rotation coupling.
\end{abstract}

\pacno{04.20.Cv}

\section{Introduction}

The nature of the interaction between intrinsic spin and acceleration has been extensively investigated in the literature from both classical and quantum perspective, in view of possible violations of the equivalence principle between inertial and gravitational masses as well as of parity and time-reversal invariance in the gravitational interaction (see, e.g., Ref. \cite{mash00} for a review).
Several studies of the Dirac equation in uniformly accelerated reference frames and gravitational fields have shown that a  {\it direct} coupling of spin with linear acceleration does not arise \cite{hehl,ryder,bcm04}. This is in agreement with the lack of observational evidence in favor of such a coupling. 
In contrast, there exists a coupling of intrinsic spin with rotation, which is independent of the inertial mass of the particle \cite{mash88}. 
For instance, in the case of photons the helicity-rotation coupling is responsible for the phenomenon of phase wrap-up, which has been tested with high accuracy via rotating GPS receivers \cite{ashby}. Furthermore, in the case of neutrinos this kind of interaction may lead to a helicity flip \cite{cai}.  

In this paper we study the behavior of massless spin-$\frac{1}{2}$ particles in the vacuum C-metric background, which represents the exterior gravitational field of a uniformly accelerating Schwarzschild black hole, under certain conditions. We refer to Ref. \cite{grifpod} for an exhaustive review of its main geometrical and physical properties as well as its historical background. 
By using spherical-like coordinates, the Dirac equation is separable into one-dimensional radial and angular parts. The radial and angular equations are coupled via the angular eigenvalue. We consider first the angular part as an eigenvalue problem, discussed here to leading order in perturbation theory, following the approach of Press and Teukolsky \cite{press}. We provide the corrections to the angular functions as well as to the associated eigenvalues to first order in a suitably defined acceleration parameter. 
The radial equation is then solved perturbatively up to a certain post-Newtonian order.
The particle absorption rate as well as the angle-averaged cross section of the associated scattering problem are also computed in the low-frequency limit.
Noticeably, both radial and angular equations can be solved exactly in terms of general Heun functions, which are but of poor practical use and only of formal utility. 
Finally, we discuss some features of the coupling between helicity and acceleration.

Units are chosen so that $G=1=c=\hbar$ and the metric signature is $-2$.

\section{Dirac equation in the C-metric}

The C-metric belongs to a class of degenerate metrics discovered by Levi-Civita \cite{LC} and takes his name from the classification of Ehlers and Kundt \cite{kundt}.
It represents the exterior field of a uniformly accelerated spherical gravitational source and it can also be thought of as a nonlinear superposition of the (flat) Rindler spacetime associated with a uniformly accelerated family of observers and the Schwarzschild solution for a static black hole \cite{kinwal,bon83}.  
The line element has been expressed using many different coordinate systems, those more suitable to study their geometrical properties \cite{Farh,dray,bicak89,pravda,hong,griffths}.
We will adopt Schwarzschild-like coordinates $x^\alpha=(t,r,\theta,\phi)$, in terms of which the line element writes as \cite{griffths}
\begin{equation}
\label{c-metric}
\fl\qquad
ds^2 = \frac{1}{\Omega^2(r,\theta)} \left( \frac{Q(r)}{r^2}\, dt^2 - \frac{r^2}{Q(r)}\, dr^2 - \frac{r^2}{P(\theta)}\, d\theta^2 - r^2P(\theta)\sin^2\theta\, d\phi^2\right),
\end{equation}
where $Q(r)=r(r-2M)(1-A^2r^2)$, $P(\theta)=1-2MA\cos\theta$ and $\Omega(r,\theta)=1-Ar\cos\theta$. 
The constants $M\ge 0$ and $A\ge 0$ denote the mass and acceleration of the source, respectively. The metric (\ref{c-metric}) then reduces to that of a Schwarzschild black hole for $A=0$ and to the flat spacetime in uniformly accelerating coordinates for $M=0$.
In this coordinate system, $r$ is constrained to the interval $2M<r<1/A$ between Schwarzschild and Rindler horizons.
Furthermore, one should require $P(\theta)>0$ to preserve metric signature, i.e., $\cos\theta<1/(2MA)$, thus excluding an entire conical region around the positive $z-$axis. 
However, if one limits the acceleration parameter so that $MA<1/2$, then $P(\theta)>0$ for all $\theta\in[0,\pi]$. Hereafter we will assume this condition to hold. 
Unavoidably, a conical singularity still exists, since $P(0)=1-2MA\not=1$.
The latter can be removed by limiting the range of allowed values of $\phi$ in the interval $\phi=[0,2\pi/P(0)]$.
No further constraint comes from the conformal factor $\Omega$, which is everywhere positive between the Schwarzschild and Rindler horizons.

Let us introduce the following Kinnersley-like null  frame \cite{kin69} 
\begin{eqnarray}
\label{NP_frame}
l=\frac{\Omega^2 r^2}{Q} \,\left(\partial_t+\frac{Q}{r^2}\partial_r\right),\quad 
n=\frac{1}{2}\,\left(\partial_t-\frac{Q}{r^2}\partial_r\right),\nonumber\\
m=\frac{\Omega}{\sqrt{2}r}\, \left(\sqrt{P}\partial_\theta + \frac{i}{\sqrt{P}\sin\theta}\partial_\phi\right).
\end{eqnarray}
By using standard  notations and conventions of the NP formalism (see e.g., Ref. \cite{chandra}) we analyze below the dynamics of Dirac particles (with special attention to the massless case). 
The wave function associated with such particles is written in terms of a pair of spinors $P^A$ and $\bar{Q}^{A'}$ (spinor indices  are denoted by capital letters and run from 0 to 1), whose components are often indicated as 
$P^0=F_1$, $P^1=F_2$, $\bar{Q}^{1'}=G_1$ and $\bar{Q}^{0'}=-G_2$, resulting in the following equations
\begin{eqnarray}
(D+\varepsilon-\rho)F_1+(\delta^*+\pi-\alpha)F_2=i\mu_*G_1,\label{dir-gen-c1}\nonumber\\
(\Delta+\mu-\gamma)F_2+(\delta+\beta-\tau)F_1=i\mu_*G_2,\label{dir-gen-c2}\nonumber\\
(D+\varepsilon^*-\rho^*)G_2-(\delta+\pi^*-\alpha^*)G_1=i\mu_*F_2,\label{dir-gen-c3}\nonumber\\
(\Delta+\mu^*-\gamma^*)G_1-(\delta^*+\beta^*-\tau^*)G_2=i\mu_*F_1\label{dir-gen-c4}\,;
\end{eqnarray}
here $\mu_*$ is the particle mass and $D\equiv l^{\mu}\partial_{\mu}$, $\Delta\equiv n^{\mu}\partial_{\mu}$, $\delta\equiv m^{\mu}\partial_{\mu}$ and $\delta^*\equiv \bar m^{\mu}\partial_{\mu}$ denote frame derivatives. The nonvanishing spin coefficients associated with the frame (\ref{NP_frame}) are given by
\begin{eqnarray}\fl\quad
\rho&=&-\frac{\Omega \left(\Omega - r\,\Omega_r \right)}{r}\,,\quad
\tau=-\pi=\frac{\sqrt {P}}{\sqrt{2}r}\, \Omega_\theta\,,\quad
\mu=\frac{Q}{2r^2}\,\frac{\rho}{\Omega^2}\,,\quad
\alpha=\frac{\sqrt{P}}{\sqrt{2}r} \Omega_\theta-\beta\,,\nonumber\\
\fl\quad
\beta&=&\frac{\sqrt {2}}{8}\,\frac{\Omega \sin\theta\, P_\theta +2\,P \cos\theta}{r\sqrt{P}\sin\theta}\,,\quad
\gamma=\frac{1}{4}\,\frac{r \Omega Q_r -2\,r Q \Omega_r-2\,Q\,\Omega}{r^3\Omega}\,,
\end{eqnarray}
where $X_r=\partial_r X$ and $X_\theta=\partial_\theta X$.
The only nonvanishing Weyl scalar is $\psi_2=-M\Omega^3/r^3$.

The existence of a time-like Killing vector $\partial_t$ and a rotational one $\partial_\phi$ in the C-metric allows us to assume a wave function with the customary dependence $e^{-i(\omega t - m \phi)}$. It is useful to define the following radial and angular differential operators as
\beq\fl\qquad
{\cal D}_n= \frac{\partial}{\partial r} - \frac{i\omega r^2}{Q} + n\frac{Q_r}{Q}\,, \qquad 
{\cal D}_n^\dag=({\cal D}_n)^*
= \frac{\partial}{\partial r} + \frac{i\omega r^2}{Q} + n\frac{Q_r}{Q}\,,
\eeq
and
\beq\fl\qquad
{\cal L}_{n} = \frac{\partial}{\partial\theta} + \frac{m\csc\theta}{P}+n\cot\theta\,,\quad
{\cal L}_{n}^\dag ={\cal L}_{n}(-m)
=\frac{\partial}{\partial\theta} - \frac{m\csc\theta}{P}+n\cot\theta\,,
\eeq
so that $D = \Omega^2{\cal D}_0$, $\Delta = -\frac{Q}{2r^2}{\cal D}_0^\dag$, $\delta = \frac{\Omega\sqrt{P}}{\sqrt{2}\,r} {\cal L}_{0}$ and $\delta^* = \frac{\Omega\sqrt{P}}{\sqrt{2}\,r} {\cal L}_{0}^\dag$. 
The following rescaling of the quantities $F_{1,2}$ and $G_{1,2}$
\beq\fl
f_1=\frac{rP^{1/4}}{\Omega}F_1,\quad f_2=\frac{\sqrt{Q}P^{1/4}}{\Omega^2}F_2,\quad 
g_1=\frac{\sqrt{Q}P^{1/4}}{\Omega^2}G_1, \quad \mbox{and}\quad g_2=\frac{rP^{1/4}}{\Omega}G_2\,,
\eeq
allows for considerable simplifications in Eqs.\ (\ref{dir-gen-c1}), which become
\begin{eqnarray}
\sqrt{Q}\,{\cal D}_0f_1 + \sqrt{\frac{P}{2}}\,{\cal L}_{\frac{1}{2}}f_2 &=& \frac{i\mu_*r\,g_1}{\Omega},\label{dir-gen-mass1}\nonumber\\[1ex]
\sqrt{Q}\,{\cal D}_0^\dag f_2 - \sqrt{2P}\,{\cal L}_{\frac{1}{2}}^\dag f_1 &=& - 2\frac{i\mu_*r\,g_2}{\Omega}, \label{dir-gen-mass2}\nonumber\\[1ex]
\sqrt{Q}\,{\cal D}_0g_2 - \sqrt{\frac{P}{2}}\,{\cal L}_{\frac{1}{2}}^\dag g_1 &=& \frac{i\mu_*r\,f_2}{\Omega},\label{dir-gen-mass3}\nonumber\\ [1ex]
\sqrt{Q}\,{\cal D}_0^\dag g_1 + \sqrt{2P}\,{\cal L}_{\frac{1}{2}}g_2 &=& -2\frac{i\mu_*r\,f_1}{\Omega}.\label{dir-gen-mass4}
\end{eqnarray}

Let us consider the case of massless particles, i.e., $\mu_*=0$.
Eqs.\ (\ref{dir-gen-mass1}) can then be solved by separation of variables. In fact, defining
\begin{equation}
\label{sep-var}
\fl\qquad
\left\{\begin{array}{l}
f_1(r,\theta) = \frac{1}{\sqrt2}R_{-\frac{1}{2}}(r)S_{-\frac{1}{2}}(\theta), \quad f_2(r,\theta) = R_{+\frac{1}{2}}(r)S_{+\frac{1}{2}}(\theta),\\[1ex]
g_1(r,\theta) = R_{+\frac{1}{2}}(r)S_{-\frac{1}{2}}(\theta), \quad g_2(r,\theta) = \frac{1}{\sqrt2}R_{-\frac{1}{2}}(r) S_{+\frac{1}{2}}(\theta),
\end{array}\right.
\end{equation}
we get  the radial  equations
\begin{eqnarray}
\label{dir-mlr2}
\sqrt{Q}\,{\cal D}_0R_{-\frac{1}{2}} = \lambda R_{\frac{1}{2}}, \qquad 
\sqrt{Q}\,{\cal D}_0^\dag R_{\frac{1}{2}} = \lambda R_{-\frac{1}{2}}, 
\end{eqnarray}
and  the angular equations
\begin{eqnarray}
\label{dir-mlt2}
\sqrt{P}\,{\cal L}_{\frac{1}{2}}S_{\frac{1}{2}} = -\lambda S_{-\frac{1}{2}}, \qquad
\sqrt{P}\,{\cal L}_{\frac{1}{2}}^\dag S_{-\frac{1}{2}} = \lambda S_{\frac{1}{2}}, 
\end{eqnarray}
where $\lambda$ is the  separation constant. 
Note that the complex conjugate angular functions $S_{\pm\frac{1}{2}}^*$ satisfy the same equations (\ref{dir-mlt2}) as $S_{\pm\frac{1}{2}}$.

\section{Angular equation}
\label{spin_weight}

By applying the differential operator ${\cal L}_{\frac{1}{2}}^\dag$ to the first equation of Eqs. (\ref{dir-mlt2}) and using the second one to eliminate $S_{-\frac{1}{2}}$, we get the following second-order differential equation for the angular function $S_{\frac{1}{2}}$
\begin{eqnarray}\fl\qquad
\label{eqSp}
0&=&\frac{\rmd^2 S_{\frac{1}{2}}}{\rmd \theta^2}+\left(\frac{P_\theta}{2P}+\cot\theta\right)\frac{\rmd S_{\frac{1}{2}}}{\rmd \theta}\nonumber\\
\fl\qquad
&&
-\frac12\left[
1-\frac{2\lambda^2}{P}+\frac{P_\theta}{2P^2\sin\theta}(2m-P\cos\theta)+\frac{(2m+P\cos\theta)^2}{2P^2\sin^2\theta}
\right]S_{\frac{1}{2}}\,.
\end{eqnarray}
The corresponding equation for $S_{-\frac{1}{2}}$ can be obtained from Eq.\ (\ref{eqSp}) simply by replacing $m\rightarrow-m$.
Both cases $s=\pm\frac12$ can be handled together by introducing the following equation \cite{bini2}
\beq
\label{eq_ang}
\frac{1}{\sin\theta}\frac{\rmd }{\rmd \theta} 
  \left(\sin\theta \frac{\rmd\, {}_{s}{\mathcal S}(\theta)}{\rmd \theta}\right)
  + V_{\rm (ang)}(\theta)\,{}_{s}{\mathcal S}(\theta)=0\ ,
\eeq
where ${}_{s}{\mathcal S}(\theta)=P^{1/4}S_{\pm\frac{1}{2}}$ and 
\begin{eqnarray}\fl\qquad
V_{\rm (ang)}(\theta)
&=&\frac{1+E-s^2}{P}-\frac{1}{P^2}\bigg\{
\frac{[s+(m-2sMA)\cos\theta]^2}{\sin^2\theta}\nonumber\\
\fl\qquad
&&
+(m+sMA)^2+(1-s^2)(1-MA\cos\theta)^2-M^2A^2
\bigg\}\ ,
\end{eqnarray}
where the separation constant has been replaced by $\lambda^2=E+s^2$.
We choose the following normalization for the angular functions
\beq
\int d\theta \frac{\sin\theta}{P}\,|{}_{s}{\mathcal S}(\theta)|^2=1\,,
\eeq
also implying
\beq
\label{norm_cond_S}
\int d\theta \frac{\sin\theta}{\sqrt{P}}\,|S_{\pm\frac{1}{2}}(\theta)|^2=1\,.
\eeq

Eq. (\ref{eq_ang}) represents a Sturm-Liouville eigenvalue problem for the separation constant $E$.
We will solve it perturbatively in the next section to first-order in the acceleration parameter $\eta\equiv2MA$, following the approach of Press and Teukolsky \cite{press}.
We will then provide an exact solution in terms of Heun functions.

\subsection{Perturbative solution}
\label{pert_ang}

Following Press and Teukolsky \cite{press}, Eq. (\ref{eq_ang}) can be written as an eigenvalue equation involving the sum of two operators, i.e.,
\beq
\label{calHdef}
({\cal H}_0+\eta{\cal H}_1){}_{s}{\mathcal S}=-E\,{}_{s}{\mathcal S}\,,
\eeq
to first-order in the acceleration parameter $\eta$, where  
\begin{eqnarray}
\label{calHdef2}
{\cal H}_0 &\equiv &\frac{1}{\sin\theta}\frac{\rmd }{\rmd \theta} 
  \left(\sin\theta \frac{\rmd}{\rmd \theta}\right) -\frac{m^2 +s^2 +2ms\cos\theta}{\sin^2\theta},\nonumber\\
{\cal H}_1 & \equiv &ms+L\cos\theta-2m\frac{s+m\cos\theta}{\sin^2\theta},
\end{eqnarray}
with $L=l(l+1)$.
Let
\beq\fl\qquad
\label{Edef}
E=L+\eta E_1\,,\qquad
{}_{s}{\mathcal S}_{lm}={}_{s}Y_{lm}+\eta\, {}_{s}\tilde{\mathcal S}_{lm}\,,\qquad
{}_{s}\tilde{\mathcal S}_{lm}=\sum_{l'}C^{l'}_{lm}\,{}_{s}Y_{l'm}\,,
\eeq
where ${}_{s}Y_{lm}(\theta)$ denote the spin-weighted spherical harmonics (SWSH), with $l=|s|,|s|+1,\ldots$ and $-l\leq m\leq l$. 
We then have
\begin{eqnarray}
\label{eqClp}
{\cal H}_0\,{}_{s}Y_{lm} &=&-L\,{}_{s}Y_{lm},\nonumber\\
C^{l'}_{lm}(L-L'){}_{s}Y_{l'm} & =&-({\cal H}_1+E_1){}_{s}Y_{lm}\,.
\end{eqnarray}
Multiplying (\ref{eqClp})$_2$ by $\sin^2\theta\,{}_{s}Y^*_{l'm}$ and integrating over the solid angle $d\Omega=\sin\theta\rmd\theta\rmd\phi$ gives
\begin{eqnarray}\fl\qquad
\label{Cdef}
C^{l'}_{lm}(L-L')\bra\sin^2\theta\ket_{l'}&=&-(E_1+ms)\bra\sin^2\theta\ket_{l,l'}-L\bra\cos\theta\sin^2\theta\ket_{l,l'}\nonumber\\
&&
+2ms\delta_{ll'}+2m^2\bra\cos\theta\ket_{l,l'}\,,
\end{eqnarray}
where the following notation has been used
\beq
\bra X\ket_{l,l'}\equiv\bra sl'm|X|slm\ket
=\int d\Omega\, {}_{s}Y^*_{l'm}X\,{}_{s}Y_{lm}.
\eeq 
$\bra X\ket_{l,l'}$ can be computed by using the relations between the SWSH, the Wigner rotation matrices and the Clebsch-Gordan coefficients (cf. \cite{press} and references therein).
Multiplying Eq. (\ref{eqClp})$_2$ by $\sin^2\theta{}_{s}Y^*_{lm}$ and integrating over the solid angle gives instead
\begin{eqnarray}\fl\qquad
\label{Cdef2}
C^{l'}_{lm}(L-L')\bra\sin^2\theta\ket_{l',l}&=&-(E_1+ms)\bra\sin^2\theta\ket_{l}-L\bra\cos\theta\sin^2\theta\ket_{l}\nonumber\\
&&
+2ms+2m^2\bra\cos\theta\ket_{l}\,.
\end{eqnarray}
If $l=l'$, one gets the following solution for $E_1$ (from Eq. (\ref{Cdef2}))
\beq\fl\qquad
\label{E1def}
E_1=-ms+\frac1{\bra\sin^2\theta\ket_{l}}[-L\bra\cos\theta\sin^2\theta\ket_{l}+2ms+2m^2\bra\cos\theta\ket_{l}]\,.
\eeq
If $l\not=l'$, one gets the following solution for $C^{l'}_{lm}$ (from Eq. (\ref{Cdef}))
\beq\fl 
\label{Cdef3}
C^{l'}_{lm}=\frac1{L-L'}\left[-(E_1+ms)\frac{\bra\sin^2\theta\ket_{l,l'}}{\bra\sin^2\theta\ket_{l'}}-L\frac{\bra\cos\theta\sin^2\theta\ket_{l,l'}}{\bra\sin^2\theta\ket_{l'}}+2m^2\frac{\bra\cos\theta\ket_{l,l'}}{\bra\sin^2\theta\ket_{l'}}\right]\,.
\eeq

\begin{figure}
\centering
\subfigure[]{\includegraphics[scale=0.3]{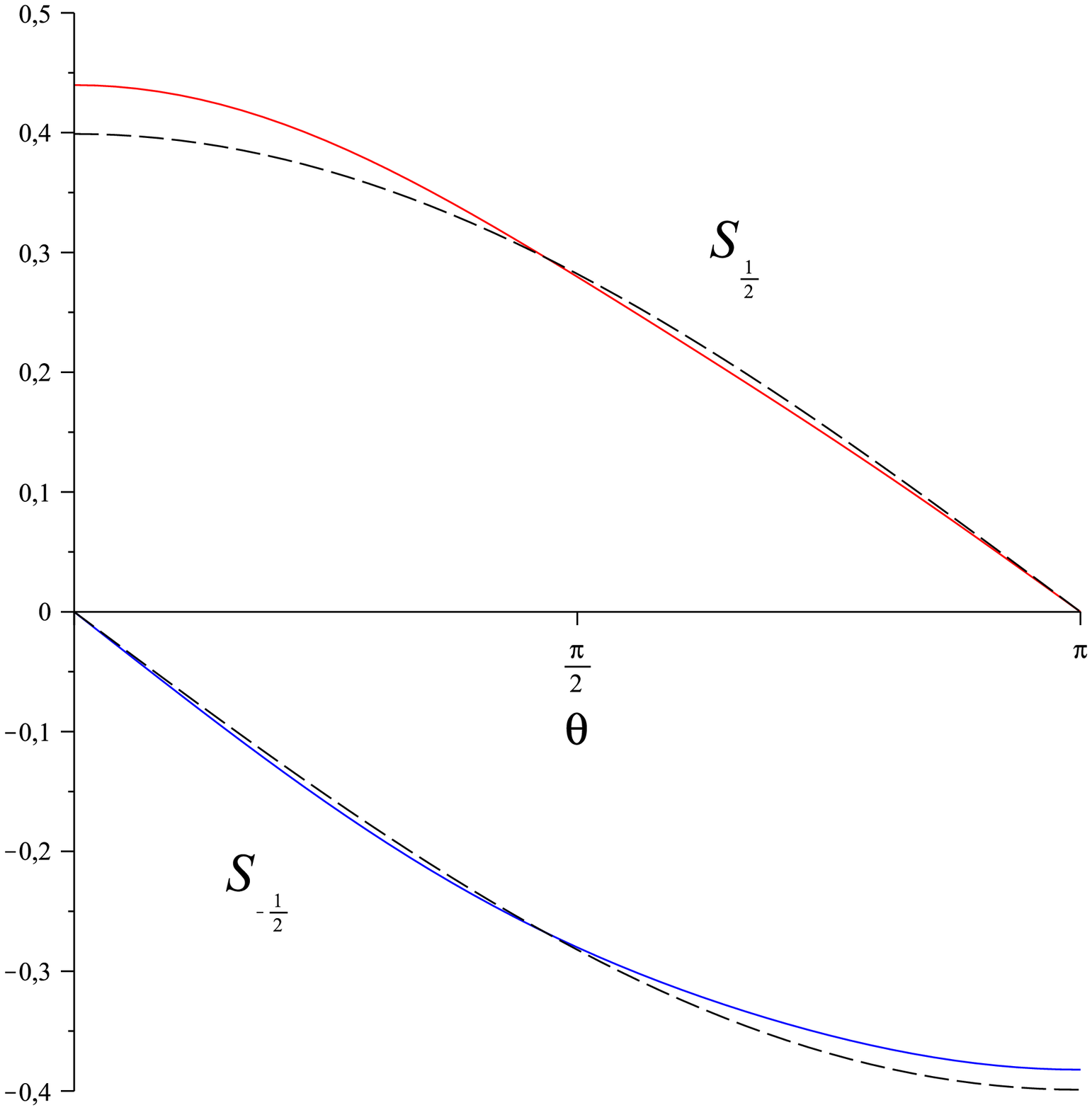}}  
\hspace{5mm}
\subfigure[]{\includegraphics[scale=0.3]{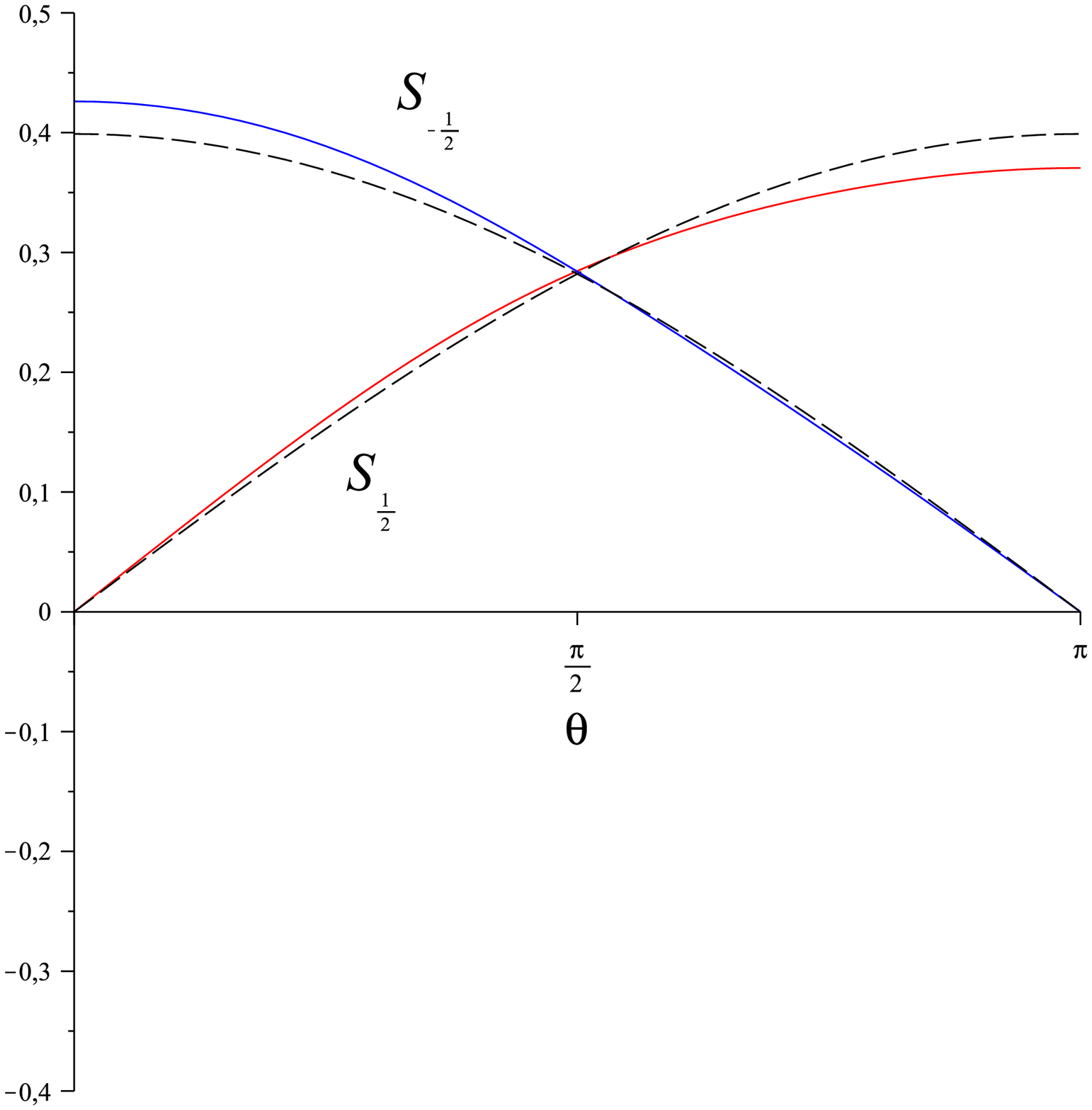}}
\caption{The behavior of the angular eigenfunctions $S_{\pm\frac{1}{2}}$ is shown for $l=1/2$ and a value $\eta=0.3$ of the acceleration parameter, as a function of the polar angle $\theta$.
Panels (a) and (b) correspond to $m=-1/2$ and $m=1/2$, respectively.
Dashed curves are the corresponding (unperturbed) SWSH ($\eta=0$).
}
\label{fig:1}
\end{figure}

\begin{figure}
\centering
\subfigure[]{\includegraphics[scale=0.3]{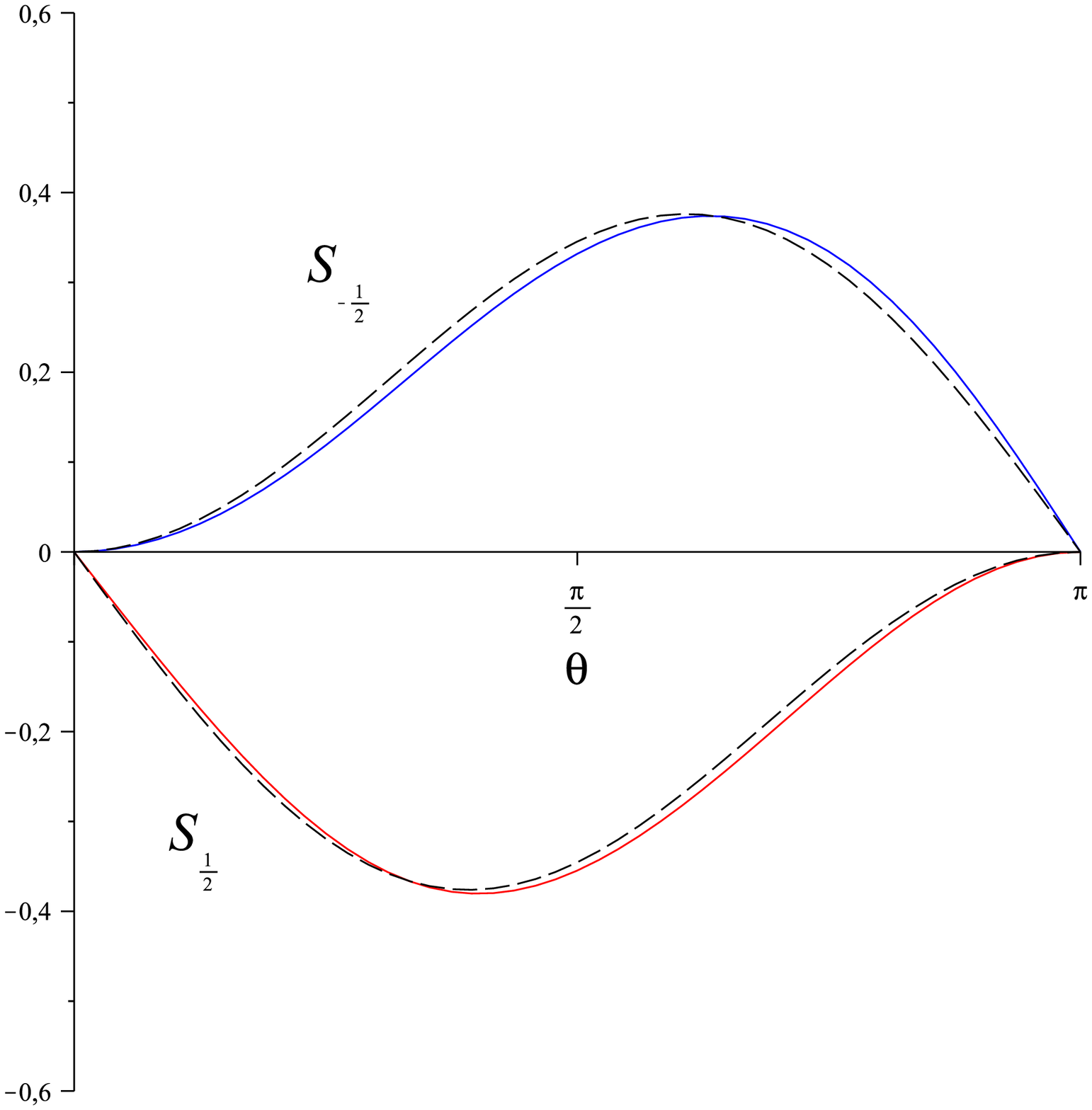}}
\hspace{5mm}
\subfigure[]{\includegraphics[scale=0.3]{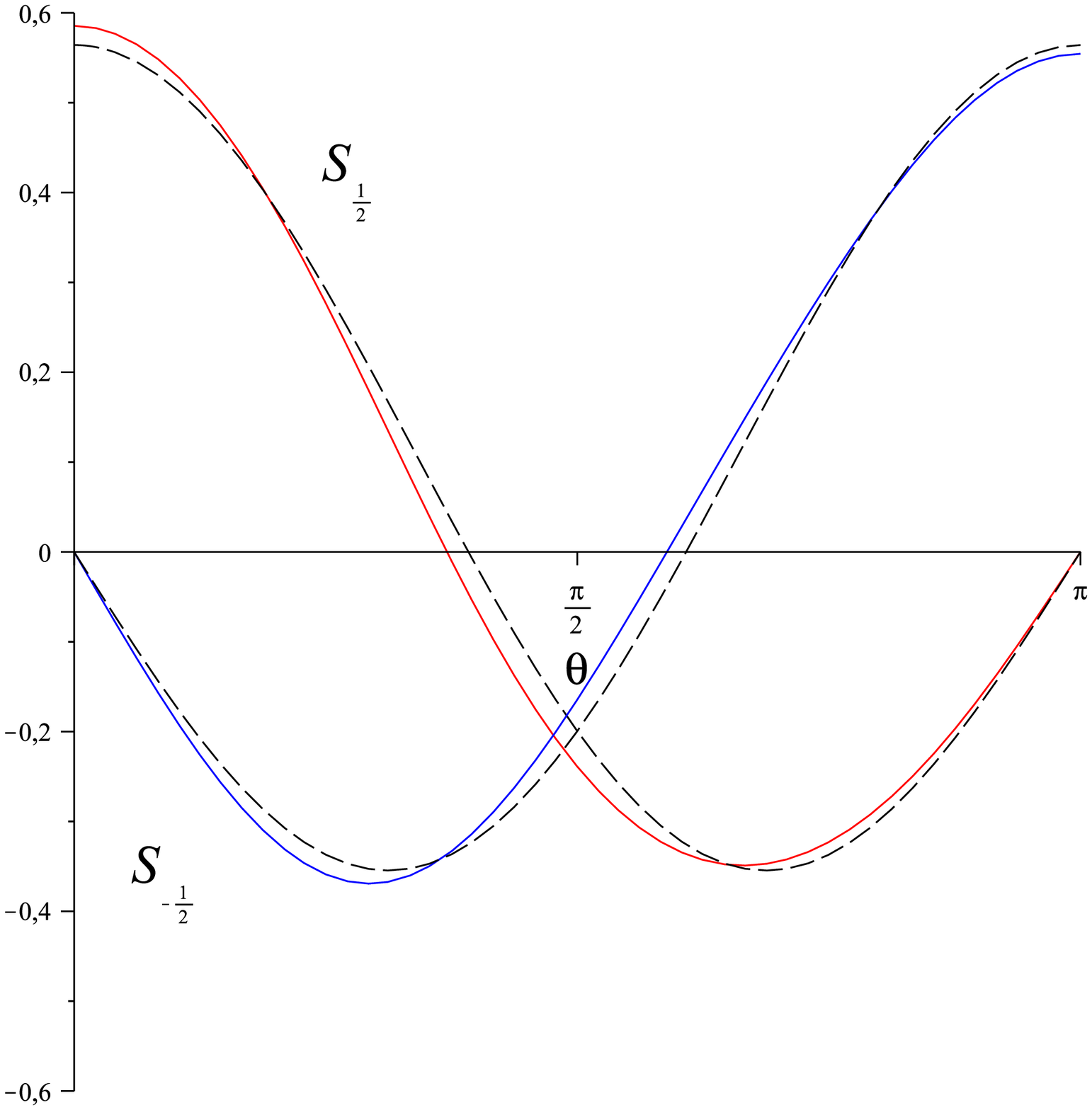}}\\[5mm] 
\subfigure[]{\includegraphics[scale=0.3]{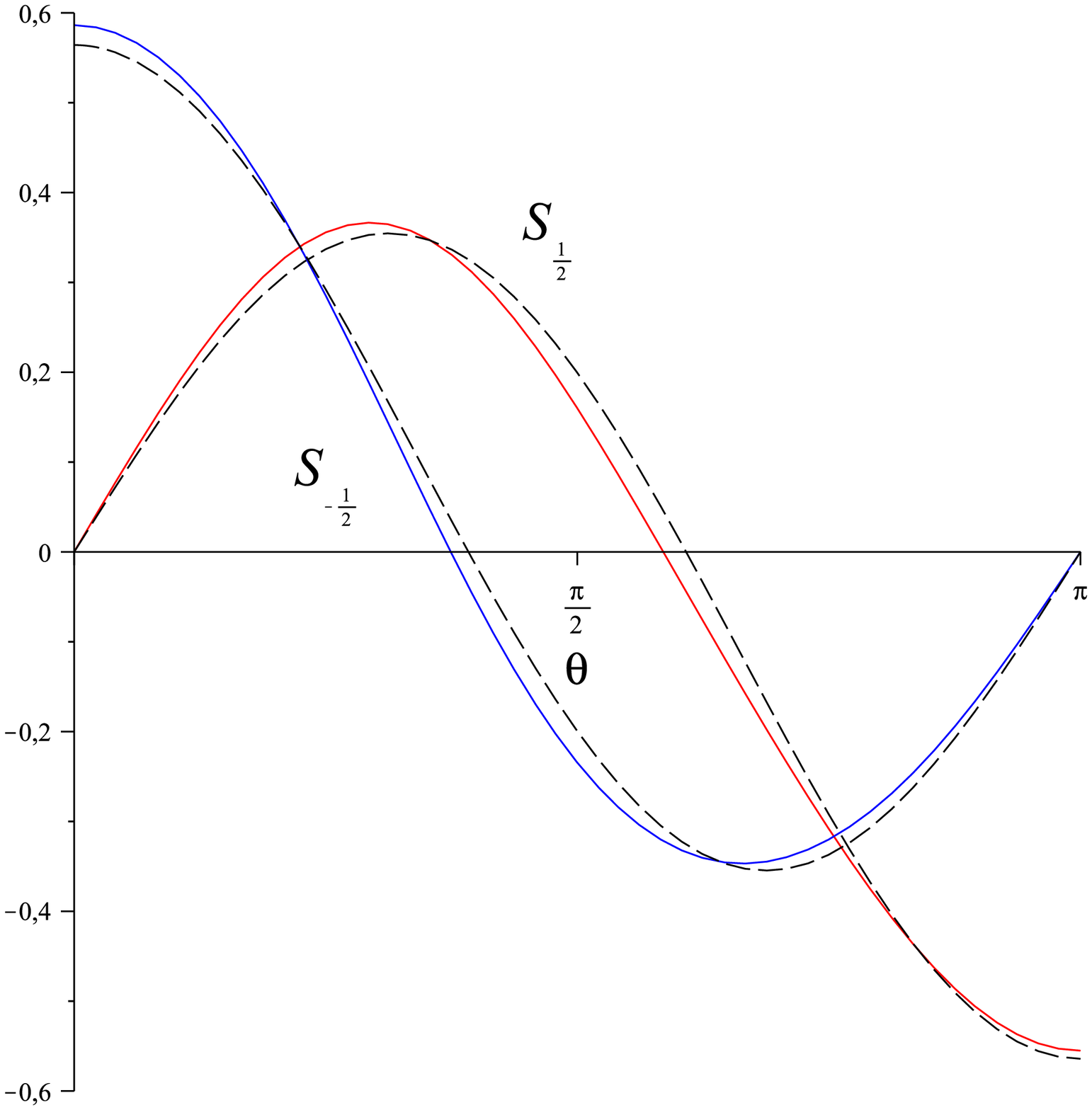}} 
\hspace{5mm}
\subfigure[]{\includegraphics[scale=0.3]{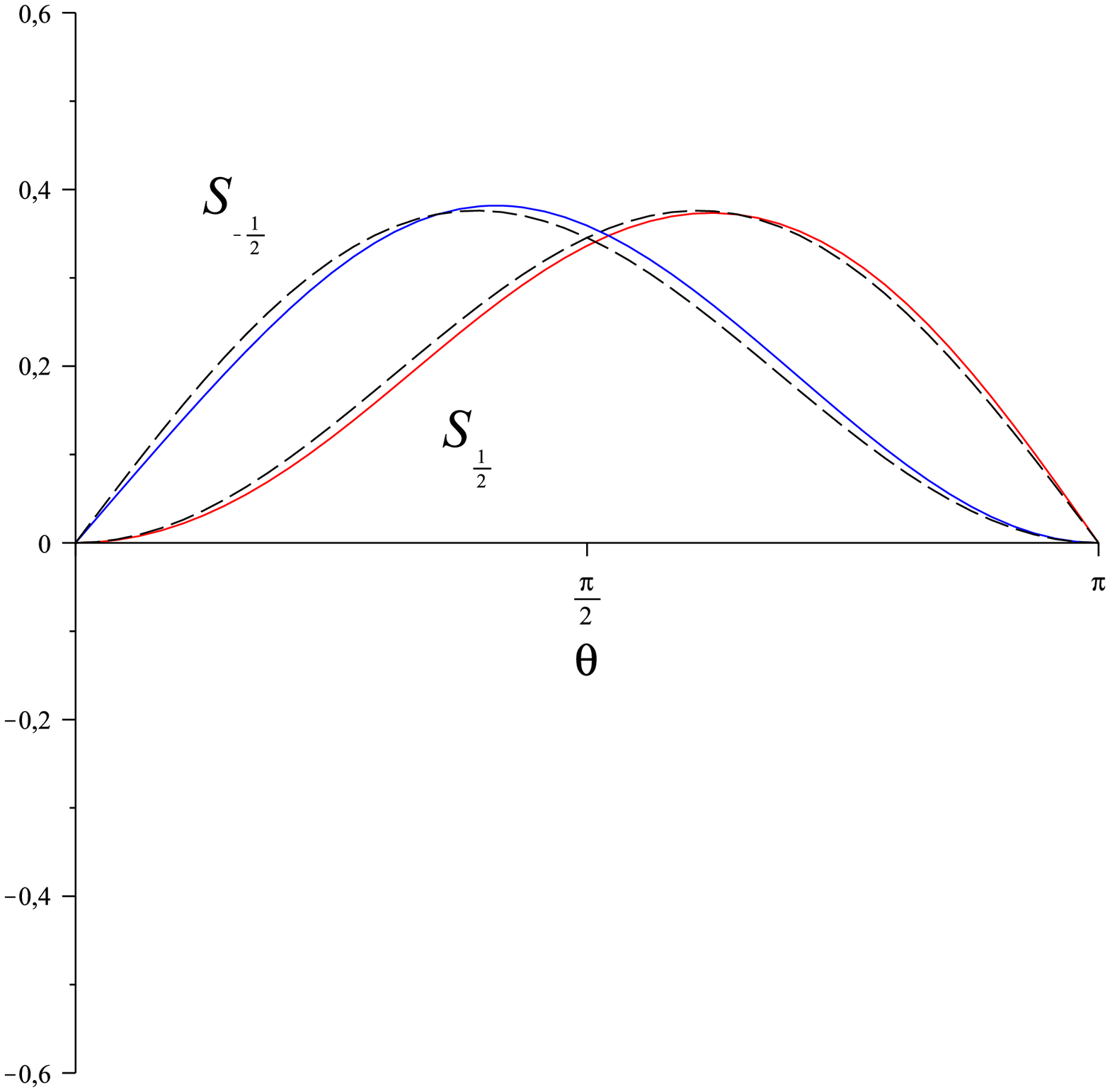}} 
\caption{The same as in Fig. \ref{fig:1}, but for the $l=3/2$ mode.
Panels (a) to (d) correspond to $m=-3/2$, $m=-1/2$, $m=1/2$ and $m=3/2$, respectively.
}
\label{fig:2}
\end{figure}

Therefore, we need to calculate the following quantities
\begin{equation}
\label{Csol}
\begin{array}{lcl}
\fl
\bra sl'm|\cos\theta|slm\ket&=&\sqrt{\frac{2l+1}{2l'+1}}\,\,\bra l 1\, m 0 | l' m\ket \bra l 1\, -s\, 0 |{l'} -s\ket,\\[2ex]
\fl
\bra sl'm|\cos^2\theta|slm\ket&=&\frac{1}{3}\delta_{ll'}+\frac{2}{3}\sqrt{\frac{2l+1}{2l'+1}}\,\,\bra l 2\, m 0 | l' m\ket \bra l 2\, -s 0 |{l'} -s\ket,\\[2ex]
\fl
\bra sl'm|\cos^3\theta|slm\ket&=&\sqrt{\frac{2l+1}{2l'+1}}\,\,\left(\frac{3}{5}\,\bra l 1\, m 0 | l' m\ket \bra l 1\, -s 0 |{l'} -s\ket\right.\\[2ex]
\fl
&&\left.+ \frac{2}{5}\,\bra l 3\, m 0 | l' m\ket \bra l 3\, -s 0 |{l'} -s\ket\right),
\end{array}
\end{equation}
where $\,\bra j_1 j_2\, m_1 m_2 | J M\ket$ are the Clebsch-Gordan coefficients, which are nonzero only if $M=m_1+m_2$, with $J=|j_1-j_2|,\ldots,j_1+j_2$ and $M=-J,\ldots,J$. 
The explicit expressions for the above quantities are shown in \ref{CGcoeffs}.
The first-order corrections to the energy eigenvalues turn out to be (see Eq. (\ref{E1sol}))
\beq
\label{E1sol_n}
E_1 = ms\frac{2[l^2(l+1)^2+5m^2]-l(l+1)(7+2m^2)}{(l-1)(l+2)[l(l+1)+m^2]}\,,
\eeq
whereas those to the coefficients $C^{l'}_{lm}$ are given by Eq.\ (\ref{Csol}).
The first-order corrections to the eigenfunctions follow directly from Eq. (\ref{Edef}).
For example, for $l=1/2,3/2,5/2$ they result in
\begin{eqnarray}\fl\quad
\label{eigencorrlist}
{}_{s}\tilde{\mathcal S}_{\frac12 m}&=&C^{\frac32}_{\frac12m}{}_{s}Y_{\frac32m}+C^{\frac52}_{\frac12m}{}_{s}Y_{\frac52m}+C^{\frac72}_{\frac12m}{}_{s}Y_{\frac72m}\,,\nonumber\\
\fl\quad
{}_{s}\tilde{\mathcal S}_{\frac32 m}&=&C^{\frac12}_{\frac32m}{}_{s}Y_{\frac12m}+C^{\frac52}_{\frac32m}{}_{s}Y_{\frac52m}+C^{\frac72}_{\frac32m}{}_{s}Y_{\frac72m}+C^{\frac92}_{\frac32m}{}_{s}Y_{\frac92m}\,,\nonumber\\
\fl\quad
{}_{s}\tilde{\mathcal S}_{\frac52 m}&=&C^{\frac12}_{\frac52m}{}_{s}Y_{\frac12m}+C^{\frac32}_{\frac52m}{}_{s}Y_{\frac32m}+C^{\frac72}_{\frac52m}{}_{s}Y_{\frac72m}+C^{\frac92}_{\frac52m}{}_{s}Y_{\frac92m}+C^{\frac{11}2}_{\frac52m}{}_{s}Y_{\frac{11}2m}\,,
\end{eqnarray}
where the coefficients are listed in Table \ref{tab:1}.
The behavior of the eigenfunctions $S_{\pm\frac{1}{2}}=P^{-1/4}{}_{\pm\frac{1}{2}}{\mathcal S}$ is shown in Figs. \ref{fig:1} and \ref{fig:2} for a given value of the acceleration parameter for the lowest modes $l=1/2$ and $l=3/2$, respectively.


\begin{table}[hbt]
\centering
\caption{The numerical values of the coefficients $C^{l'}_{lm}$ needed to compute the first-order corrections (\ref{eigencorrlist}) to the angular eigenfunctions are listed below.
}
\begin{tabular}{|c|c|c|c|c|c|c|c|}
\hline
\multicolumn{2}{|c|}{}&\multicolumn{6}{|c|}{$C^{l'}_{lm}$}\\
\hline
$l$&$m$&$l'=\frac12$ &$l'=\frac32$ &$l'=\frac52$ &$l'=\frac72$ &$l'=\frac92$ &$l'=\frac{11}2$\\
\hline
   & $-\frac12$ & $-$ & $\frac{\sqrt{2}}{80}$ & $\frac{23\sqrt{3}s}{1080}$ & $-\frac{9}{800}$ & $-$ & $-$\\
$\frac12$&&&&&&&\\
   & $\frac12$ & $-$ & $\frac{\sqrt{2}}{80}$ & $-\frac{23\sqrt{3}s}{1080}$ & $-\frac{9}{800}$ & $-$ & $-$\\
\hline
   &$-\frac32$& $0$ & $-$ & $-\frac{39}{154}$ & $\frac{\sqrt{10}s}{1680}$ & $-\frac{11\sqrt{5}}{882}$ & $-$\\
   &$-\frac12$& $-\frac{53\sqrt{2}}{280}$ & $-$ & $\frac{71\sqrt{6}}{1512}$ & $-\frac{39\sqrt{2}s}{8960}$ & $-\frac{33\sqrt{10}}{2450}$ & $-$\\
$\frac32$&&&&&&&\\
   &$\frac12$& $-\frac{53\sqrt{2}}{280}$ & $-$ & $\frac{71\sqrt{6}}{1512}$ & $\frac{39\sqrt{2}s}{8960}$ & $-\frac{33\sqrt{10}}{2450}$ & $-$\\
   &$\frac32$& $0$ & $-$ & $-\frac{39}{154}$ & $-\frac{\sqrt{10}s}{1680}$ & $-\frac{11\sqrt{5}}{882}$ & $-$\\
\hline
   &$-\frac52$& $0$ & $0$ & $-$ & $-\frac{145\sqrt{6}}{792}$ & $-\frac{115\sqrt{21}s}{31248}$ & $-\frac{65\sqrt{14}}{6804}$\\
	 &$-\frac32$& $0$ & $-\frac{43}{462}$ & $-$ & $-\frac{113\sqrt{10}}{3960}$ & $\frac{5\sqrt{5}s}{3024}$ & $-\frac{65\sqrt{35}}{6156}$ \\
   &$-\frac12$& $\frac{169\sqrt{3}s}{1080}$ & $-\frac{467\sqrt{6}}{3024}$ & $-$ & $\frac{323\sqrt{3}}{2160}$ & $\frac{17\sqrt{15}s}{7560}$ & $-\frac{325\sqrt{2}}{5832}$\\
$\frac52$&&&&&&&\\
   &$\frac12$& $-\frac{169\sqrt{3}s}{1080}$ & $-\frac{467\sqrt{6}}{3024}$ & $-$ & $\frac{323\sqrt{3}}{2160}$ & $-\frac{17\sqrt{15}s}{7560}$ & $-\frac{325\sqrt{2}}{5832}$\\
	 &$\frac32$& $0$ & $-\frac{43}{462}$ & $-$ & $-\frac{113\sqrt{10}}{3960}$ & $-\frac{5\sqrt{5}s}{3024}$ & $-\frac{65\sqrt{35}}{6156}$\\
   &$\frac52$& $0$ & $0$ & $-$ & $-\frac{145\sqrt{6}}{792}$ & $\frac{115\sqrt{21}s}{31248}$ & $-\frac{65\sqrt{14}}{6804}$\\
\hline
\end{tabular}
\label{tab:1}
\end{table}


\subsection{Exact solution}

Eq.\ (\ref{eq_ang}) can also be solved exactly in terms of Heun functions.
In fact, suitably rescaling the function ${}_{s}{\mathcal S}$ as
\beq\fl\quad
{}_{s}{\mathcal S}=yP^kz^{k_+}(z-1)^{k_-}\,, \qquad 
2k=s+1+\frac{2m\eta}{\eta^2-1}\,,\qquad
2k_\pm=s-\frac{m\eta}{\eta\pm1}\,,
\eeq
in terms of the new variable $z=\cos^2\frac\theta2$, Eq. (\ref{eq_ang}) becomes
\beq
\label{heun}
\frac{d^{2}y}{dz^2}+\left(\frac{\gamma}{z}
+\frac{\delta}{z-1}+\frac{\epsilon}{z-a}\right)\frac{dy}{dz}
+\frac{\alpha\beta-q}{z(z-1)(z-a)}y=0\,,
\eeq
which is a General Heun equation \cite{ronveaux} in standard form with solution
\begin{eqnarray}\fl\quad
\label{heunGsol}
y&=&c_1{\rm HeunG}[a,q,\alpha,\beta,\gamma,\delta;z]\nonumber\\
\fl\quad
&&
+c_2z^{1-\gamma}{\rm HeunG}[a,q+(1-\gamma)(a\delta+\epsilon),\beta+1-\gamma,\alpha+1-\gamma,2-\gamma,\delta;z]\,,
\end{eqnarray}
and parameters
\begin{eqnarray}\fl\quad
2\eta a&=&1+\eta\,,\qquad
2\eta q=(1+2\eta)s^2+(1+3\eta)s+\eta-E\,,\nonumber\\
\fl\quad
\alpha&=&1+s\,,\qquad
\beta=1+2s\,,\qquad
\gamma=1+2k_+\,,\qquad
\delta=1+2k_-\,,
\end{eqnarray}
with $\gamma+\delta+\epsilon=\alpha+\beta+1$ and $a\not=0,1$.
This equation has four regular singular points at $z_i=\{0, 1, a,\infty\}$.
In fact, the coefficient of $y'$ in Eq. (\ref{heun}) has a singular part diverging as $A_i(z-z_i)^{-1}$, whereas that of $y$ behaves as $B_i(z-z_i)^{-2}+C_i(z-z_i)^{-1}$, with at least one of the coefficients $A_i$, $B_i$ and $C_i$ nonvanishing.
Furthermore, the General Heun function is such that 
\beq\fl\quad
{\rm HeunG}[a,q,\alpha,\beta,\gamma,\delta;0]=1\,,\qquad
\frac{\rmd}{\rmd z}{\rm HeunG}[a,q,\alpha,\beta,\gamma,\delta;0]=\frac{q}{\gamma a}\,,
\eeq
and its expansion around $z=0$ starts with 
\begin{eqnarray}\fl\qquad
&&{\rm HeunG}[a,q,\alpha,\beta,\gamma,\delta;z]=
1+\frac{q}{\gamma a}z\nonumber\\
\fl\qquad
&&\qquad
+\frac{q+qa\delta+q\gamma a-q\delta+q\alpha+q\beta+q^2-\alpha\beta\gamma a}{\gamma(\gamma+1)a^2}z^2+O(z^3)\,.
\end{eqnarray}

\section{The radial equation}
\label{rad_schr_cmet}

Let us consider the radial part of the Dirac equation. Combining Eqs. (\ref{dir-mlr2}), we get the following second order equations for the radial function $R_{-\frac{1}{2}}$ 
\begin{equation}
(Q{\cal D}_{\frac{1}{2}}^\dag{\cal D}_0-\lambda^2)R_{-\frac{1}{2}} = 0, 
\end{equation}
which reads
\beq\fl\qquad
\frac{\rmd^2 R_{-\frac{1}{2}}}{\rmd r^2}+\frac{Q_r}{2Q}\frac{\rmd R_{-\frac{1}{2}}}{\rmd r}
-\frac1Q\left[
\lambda^2-\frac{\omega^2r^4}{Q}+2i\omega r\left(1-\frac{rQ_r}{4Q}\right)
\right]R_{-\frac{1}{2}}=0\,,
\eeq
whereas $R_{\frac{1}{2}}$ satisfies the complex conjugate equation.
Both cases $s=\pm\frac12$ can be handled together by introducing the following equation \cite{bini2}
\beq
\label{radeq}
Q^{-s} \frac{\rmd}{\rmd r}\left(Q^{s+1}\frac{\rmd {}_{s}{\mathcal R}(r)}{\rmd r} \right)
+V_{\rm (rad)}(r){}_{s}{\mathcal R}(r)=0\ ,
\eeq
with
\begin{eqnarray}\fl\qquad
V_{\rm (rad)}(r) &=&-2rA^2(r-M)(1+s)(1+2s)+\frac{\omega^2r^4}{Q} \nonumber \\
\fl\qquad
&& -2is\omega r \left(
\frac{M}{r-2M}-\frac{1}{1-A^2r^2}
\right)-\lambda^2+s(1+2s)\,,
\end{eqnarray}
and the new radial function ${}_{s}{\mathcal R}$ is such that ${}_{-\frac12}{\mathcal R}=R_{-\frac{1}{2}}$ and $\sqrt{Q}\,{}_{\frac12}{\mathcal R}=R_{\frac{1}{2}}$.

\subsection{Perturbative solution}

Let us consider first perturbative solutions to the radial equation (\ref{radeq}) for small values of the acceleration parameter $\eta=2MA$.
Since $A$ enters this equation quadratically, to first-order in $\eta$ the radial equation maintains the same form as in the Schwarzschild case ($\eta=0$), differently from the angular equation.
Nevertheless, a dependence on the acceleration parameter still remains through the separation constant, which is such that $\lambda^2=E+s^2$, and the energy eigenvalues $E=L+\eta E_1$ are affected to first-order in $\eta$ according to Eq. (\ref{E1sol}). 

We list below the first few terms of the corresponding PN solution ($M\sim\epsilon^2$, $A\sim\epsilon^{-2}$ and $\omega\sim\epsilon$, where $\epsilon\equiv1/c$ is the inverse of the speed of light) to first-order in $\eta$.
The two independent solutions are
\beq\fl\qquad
\label{Rinupdefs}
{}_{s}{\mathcal R}_{\rm (in)}={}_{s}{\mathcal R}_{\rm (in)}^{(0)}+\eta E_1\,{}_{s}{\mathcal R}_{\rm (in)}^{(1)}\,,\qquad
{}_{s}{\mathcal R}_{\rm (up)}={}_{s}{\mathcal R}_{\rm (up)}^{(0)}+\eta E_1\,{}_{s}{\mathcal R}_{\rm (up)}^{(1)}\,,
\eeq
where the subscripts \lq\lq in'' and \lq\lq up'' refer to the regularity for small and large values of $r$, respectively.
We find
\begin{eqnarray}\fl\qquad
{}_{s}{\mathcal R}_{\omega lm}{}_{\rm (in)}^{(0)}&=&r^{l-s}\left\{
1-\frac{is\omega r}{l+1}\epsilon
-\left[(l-s)\frac{M}{r}+\frac{\omega^2r^2}{4(l+1)}\right]\epsilon^2
+\frac{is\omega^3r^3}{4(l+1)(l+2)}\epsilon^3\right.\nonumber\\
\fl\qquad
&&\left.
+\frac14\left[(l-1)(2l-1-4s)\frac{M^2}{r^2}+\frac{(l+1)(l-7)-ls}{(l+1)^2}M\omega^2r
\right.\right.\nonumber\\
\fl\qquad
&&\left.\left.
+\frac{\omega^4r^4}{8(l+1)(l+2)}\right]\epsilon^4
+O(\epsilon^5)
\right\}
\,,\nonumber\\
\fl\qquad
{}_{s}{\mathcal R}_{\omega lm}{}_{\rm (up)}^{(0)}&=&{}_{s}{\mathcal R}_{\omega lm}{}_{\rm (in)}^{(0)}\big\vert_{l\to-l-1}\,,
\end{eqnarray}
and
\begin{eqnarray}\fl
{}_{s}{\mathcal R}_{\omega lm}{}_{\rm (in)}^{(1)}&=&\frac{r^{l-s}}{(2l+1)^2}\left\{
(2l+1)\ln\left(\frac{r}{r_0}\right) -1\right.\nonumber\\
\fl
&&
-\frac{is\omega r}{(l+1)^2}\left[(l+1)(2l+1)\ln\left(\frac{r}{r_0}\right)-3l-2\right]\epsilon\nonumber\\
\fl
&&
-\frac{2l-1}{4(l-s)(l-s-1)}\left[\frac{M}{r}[(2l+1)(2l^2-l-6ls+1+s)\ln\left(\frac{r}{r_0}\right)\right.\nonumber\\
\fl
&&
-(2l+5)s+2l^2+l-2]\nonumber\\
\fl
&&\left.
+\frac{\omega^2r^2}{4(l+1)^2}(2l-1-4s)[(l+1)(2l+1)\ln\left(\frac{r}{r_0}\right)-3l-2]
\right]\epsilon^2\nonumber\\
\fl
&&
+\frac{is\omega^3r^3(2l-1)(2l-1-4s)}{16(l+1)^2(l+2)^2(l-s)(l-s-1)}\left[(l+1)(l+2)(2l+1)\ln\left(\frac{r}{r_0}\right)\right.\nonumber\\
\fl 
&&\left.
-5l^2-11l-5\right]\epsilon^3\nonumber\\
\fl
&&
+\frac{2l-1}{16(l-s)(l-s-1)}\left[\frac{M^2}{r^2}[(l-1)(2l+1)(4l^2-16ls-4l+8s+5)\ln\left(\frac{r}{r_0}\right)\right.\nonumber\\
\fl
&&
+(-16l+24-32l^2)s-5l+12l^3+12-4l^2]\nonumber\\
\fl
&&
+\frac{M\omega^2r}{(l+1)^3}[(l+1)(2l+1)(2l^3-13l^2-6l^2s-7l+25ls+28s+7)\ln\left(\frac{r}{r_0}\right)\nonumber\\
\fl
&&
+10l^3s-87l^2s+50l^2-150ls-7l-59s-14-2l^4+43l^3]\nonumber\\
\fl
&&
+\frac{\omega^4r^4}{8(l+1)^2(l+2)^2}(2l-1-4s)[(l+1)(l+2)(2l+1)\ln\left(\frac{r}{r_0}\right)\nonumber\\
\fl
&&\left.\left.
-5l^2-11l-5]
\right]\epsilon^4
+O(\epsilon^5)
\right\}
\,,\nonumber\\
\fl
{}_{s}{\mathcal R}_{\omega lm}{}_{\rm (up)}^{(1)}&=&{}_{s}{\mathcal R}_{\omega lm}{}_{\rm (in)}^{(1)}\big\vert_{l\to-l-1}\,,
\end{eqnarray}
where $r_0$ is an arbitrary scale factor.
The constant Wronskian 
\beq
W=Q^{s+1}\left[{}_{s}{\mathcal R}_{\omega lm}{}_{\rm (in)}\frac{\rmd}{\rmd r}{}_{s}{\mathcal R}{}_{\omega lm}{}_{\rm (up)}
-{}_{s}{\mathcal R}_{\omega lm}{}_{\rm (up)}\frac{\rmd}{\rmd r}{}_{s}{\mathcal R}{}_{\omega lm}{}_{\rm (in)}\right]
\eeq
is given by 
\begin{eqnarray}\fl\qquad
W&=&-2l-1+isM\omega\frac{[2l(l+1)-7s-4](2l-1)(2l+3)(2l+1)^3}{16l(l+1+s)(l-s)(l+1)(l-s-1)(l+2+s)}\nonumber\\
\fl\qquad
&&\times
\left[
1-\eta E_1\frac{4sl(l+1)-7s-4}{(2l+1)^2l(l+1)[2l(l+1)-7s-4]}
\right]\epsilon^3+O(\epsilon^5,\eta^2)\,.
\end{eqnarray}

\subsection{Exact solution}

As in the angular case, the radial equation (\ref{radeq}) can be solved exactly in terms of General Heun functions.
In fact, suitably rescaling the function ${}_{s}{\mathcal R}$ as
\beq 
{}_{s}{\mathcal R}=wQ^{-s}(r-2M)^{k_1}(1+Ar)^{k_2}(Ar-1)^{k_3}\,,
\eeq
with 
\beq\fl\qquad
k_1=-\frac{2i\omega M}{\eta^2-1}\,,\qquad
k_2=s+\frac{i\omega M}{\eta(\eta+1)}\,,\qquad
k_3=-1+\frac{i\omega M}{\eta(\eta-1)}\,,
\eeq
in terms of the new variable 
\beq
z=\frac{r}{2M}\frac{1-2AM}{1-Ar}\,,
\eeq 
interpolating between $z=1$ ($r=2M$) and $z\to\infty$ ($r=1/A$), Eq. (\ref{radeq}) can be cast in the form of a General Heun equation (\ref{heun}) with solution $w=w(z)$ given by Eq. (\ref{heunGsol}) with parameters
\begin{eqnarray}\fl\quad
2\eta a&=&\eta-1\,,\qquad
2\eta q=\eta(1-s^2)+E\,,\nonumber\\
\fl\quad
\alpha&=&1+s\,,\qquad
\beta=-(1+2k_3)\,,\qquad
\gamma=1-s\,,\qquad
\delta=1-s+2k_1\,.
\end{eqnarray}

\subsection{Asymptotics}

In order to study the asymptotic behavior of the radial functions at the Schwarzschild and Rindler horizons it is useful to introduce the tortoise-like coordinate $r_*$ defined by ${dr_*}/{dr}={r^2}/{Q}$, i.e.,
\begin{eqnarray}
\label{rstardef}
r_*&=&\frac1{A(1+2AM)}\ln\sqrt{1+Ar}
-\frac1{A(1-2AM)}\ln\sqrt{1-Ar}\nonumber\\
&&
+\frac{2M}{1-4M^2A^2}\ln\left(\frac{r}{2M}-1\right)
+{\rm const}\,.
\end{eqnarray}
It is a single-valued function of $r$, because $dr/dr_*$ is always positive, and is such that $r_*\rightarrow-\infty$ when $r\rightarrow2M$, and $r_*\rightarrow+\infty$ when $r\rightarrow1/A$.

By introducing the scaling ${}_{s}{\mathcal R}=r^{-1}Q^{-\frac s2}H$, in terms of the new variable $r_*$ the radial equation (\ref{radeq}) can be transformed into the one-dimensional Schr\"odinger-like equation
\beq
\label{newradi}
\frac{\rmd ^2 }{\rmd r_*^2}H(r)+\tilde V H(r)=0\ ,
\eeq
with potential
\beq
\tilde V= \frac{Q}{r^4}\left[V_{\rm (rad)}+\frac{2Q}{r^2}-\frac{Q_r}{r}-\frac{s}{2}\left(Q_{rr}+\frac{s}{2}\frac{Q_r^2}{Q}\right)\right]\,.
\eeq
The asymptotic form of the radial equation as $r\to 1/A \, (r_{*}\to\infty)$ is
\beq
\frac{\rmd ^2 }{\rmd r_*^2}H(r)+\left[\omega+is\kappa_A\right]^2 H(r)=0\,,
\eeq
where $\kappa_A=A(1-2MA)$ is the value of the surface gravity at the Rindler horizon. 
For small values of $A$ we then find
\beq
\left[\omega+is\kappa_A\right]^2 = \omega^2+ 2iAs\omega+O(A^2)\,.
\eeq
The solution is then given by
\beq
H\sim e^{\pm i[\omega+is\kappa_A]r_*}
\sim Q^{\mp s/2}e^{\pm i\omega r_*}\,,
\eeq
implying that ${}_{s}{\mathcal R}\sim Q^{-s}e^{i\omega r_*}$ and ${}_{s}{\mathcal R}\sim e^{-i\omega r_*}$ for outgoing and ingoing waves, respectively.

On the other hand  close to the Schwarzschild horizon $r\to 2M\, (r_{*}\to-\infty)$, the  asymptotic form of the radial equation becomes
\beq
\frac{\rmd ^2 }{\rmd r_*^2}H(r)+\left[\omega-is\kappa_+\right]^2 H(r)=0\,,
\eeq
where $\kappa_+=(1-4M^2A^2)/(4M)$ is the value of the surface gravity at the Schwarzschild horizon. 
For small values of $A$ we then find
\beq
\left[\omega-is\kappa_+\right]^2 = \left(\omega-is\kappa_+\right)^2+O(A^2)\,.
\eeq
Therefore, the asymptotic solution is given by
\beq
H\sim e^{\pm i[\omega-is\kappa_+]r_*}
\sim Q^{\pm s/2}e^{\pm i\omega r_*}\,,
\eeq
implying that ${}_{s}{\mathcal R}\sim e^{i\omega r_*}$ (outgoing waves) and ${}_{s}{\mathcal R}\sim Q^{-s}e^{-i\omega r_*}$ (ingoing waves).  

The radial equation (\ref{radeq}) is associated with a one-dimensional scattering problem, once suitable asymptotic boundary conditions are imposed (see Ref. \cite{press}).
An incident wave traveling towards the Schwarzschild horizon with a given amplitude will be partially reflected by the potential barrier (reaching then the Rindler horizon), and partially transmitted across the black hole horizon. This situation is realized by the conditions
\beq
\label{scatt}
{}_{s}{\mathcal R}\to\left\{
\begin{array}{ll}
TQ^{-s}e^{-i\omega r_*}\,,& r\to 2M\\[3ex]
Q^{-s}e^{-i\omega r_*}+Re^{i\omega r_*}\,,& r\to\displaystyle\frac1A\\
\end{array}
\right.
\,,
\eeq
where $R$ and $T$ are reflection and transmission coefficients, respectively.

\section{Dirac current for massless particles}

Let us study now the Dirac current for massless particles, i.e.,
\beq
J^{\mu}=\sqrt{2}\,\sigma^{\mu}_{AB'}(P^A\bar P^{B'}+Q^A\bar Q^{B'})\,,
\eeq
where $\sigma^{\mu}_{AB'}$ are the generalized Pauli matrices defined as
\begin{equation}
\sigma^{\mu}_{AB'}=\frac{1}{\sqrt{2}}
\left|\begin{array}{cc}
l^\mu&m^\mu\\
\bar m^\mu&n^\mu
\end{array}\right|.
\end{equation}
In terms of the rescaled functions $f_i$ and $g_i$ it reads 
\begin{eqnarray}
\label{gen-comp-curr}
J&=&\frac{\Omega^2}{r^2\sqrt{P}}\left[
(|f_1|^2+|g_2|^2) l + (|f_2|^2+|g_1|^2) n + (f_1f_2^{*}-g_2g_1^{*}) m\right.\nonumber\\
&&\left.
+ (f_2f_1^{*}-g_1g_2^{*}) {\bar m}
\right]\,.
\end{eqnarray}
and is conserved, i.e., $\nabla_\mu J^\mu=0$.
One can then define the time rate of the number of particles entering a $r=$ const. hypersurface as
\begin{equation}
\label{cons-net-curr}
\frac{\partial N}{\partial t}=-\int \sqrt{-g}J^r\rmd\theta\rmd\phi\,,
\end{equation}
where $g$ is the determinant of the spacetime metric, so that $\sqrt{-g}=r^2\sin\theta/\Omega^4$.

Substituting the assumption (\ref{sep-var}) into Eq.\ (\ref{gen-comp-curr}) we find
\begin{equation}
\label{mles-comp-curr}
\begin{array}{lcl}
J^t&=&\frac{\Omega^4}{2Q\sqrt{P}}\,(|R_{-\frac{1}{2}}|^2 + |R_{\frac{1}{2}}|^2)(|S_{\frac{1}{2}}|^2 + |S_{-\frac{1}{2}}|^2)\,,\\[1ex]
J^r&=&\frac{\Omega^4}{2r^2\sqrt{P}}\,(|R_{-\frac{1}{2}}|^2 - |R_{\frac{1}{2}}|^2)(|S_{\frac{1}{2}}|^2 + |S_{-\frac{1}{2}}|^2),\\[1ex]
J^\theta&=&\frac{2\,\Omega^4}{r^2\sqrt{Q}}\, \mbox{Im}\left(R_{\frac{1}{2}} R_{-\frac{1}{2}}^*\right)\,\mbox{Im}\left(S_{-\frac{1}{2}}S_{\frac{1}{2}}^*\right) ,\\[1ex]
J^\phi&=&-\frac{2\,\Omega^4}{r^2\sqrt{Q}P\sin\theta} \, \mbox{Re}\left(R_{\frac{1}{2}} R_{-\frac{1}{2}}^*\right)\,\mbox{Im}\left(S_{-\frac{1}{2}}S_{\frac{1}{2}}^*\right)\,.
\end{array}
\end{equation}
The angular equations (\ref{dir-mlt2}) imply
\begin{equation}
\label{eq-im-S}
\sin\theta\,{\rm Im}(S_{-\frac{1}{2}}S^*_{\frac{1}{2}})= {\rm const.}\,,
\end{equation}
which we set to zero.
Similarly, the radial equations (\ref{dir-mlr2}) yield
\begin{equation}
\label{eq-re-im-R1}
|R_{\frac{1}{2}}|^2 - |R_{-\frac{1}{2}}|^2 = C\,,
\end{equation}
where $C$ is an integration constant.
Therefore, the Dirac current turn out to be given by
\begin{equation}
\label{mles-comp-curr2}
J=\frac{\Omega^4}{2Q\sqrt{P}}\left(|S_{\frac{1}{2}}|^2+|S_{-\frac{1}{2}}|^2\right)(|R_{\frac{1}{2}}|^2 l + |R_{-\frac{1}{2}}|^2 n)
\equiv J_+ + J_-\,,
\end{equation}
so that $J^{\mu}J_{\mu}=2J_+\cdot J_-\geq0$, whence $J$ is in general a time like vector. 
Since the only non-trivial spatial component of $J^{\mu}$ is $J^r$, we analyze the particle flux through spherical surfaces. If we assume the normalization (\ref{norm_cond_S}) of the angular functions, then the conserved net current of particles (\ref{cons-net-curr}) becomes
\beq
\frac1{2\pi}\frac{\partial N}{\partial t}=C\,.
\eeq
Turning then to the scattering problem discussed in the previous section, one can calculate the rate at which particles enter the Schwarzschild horizon per unit time, i.e.,
\beq
\frac1{2\pi}\left(\frac{\partial N}{\partial t}\right)_{r\to2M}=\left(|R_{\frac{1}{2}}|^2 - |R_{-\frac{1}{2}}|^2\right)_{r\to2M}\,.
\eeq
The asymptotic behavior (\ref{scatt}) of ${}_{s}{\mathcal R}$ for $r\to2M$ implies 
\beq
|R_{\frac{1}{2}}|^2\sim|T|^2\,,\qquad
|R_{-\frac{1}{2}}|^2\sim Q|T|^2\to0\,,
\eeq
so that 
\beq
\frac1{2\pi}\left(\frac{\partial N}{\partial t}\right)_{r\to2M}=|T|^2\,,
\eeq
which is always positive.
Therefore, superradiance cannot occur in this case \cite{bini2,tim}. 
The net current of particles crossing the Rindler horizon is instead given by
\beq
\frac1{2\pi}\left(\frac{\partial N}{\partial t}\right)_{r\to1/A}=1-|R|^2\,.
\eeq

Let us evaluate the absorption rate of particles from the Schwarzschild horizon by using the perturbative solution of the previous section. For vanishing acceleration the transmission coefficient per unit amplitude of the incident wave must reduce to that computed by Page \cite{page76} in the case of a Schwarzschild black hole, i.e., $\Gamma_{\frac12}^{\rm schw}=M^2\omega^2$, to the leading order approximation (i.e., for $M\omega\ll1$, which is the range of validity of our approximate solution, and $l=1/2$).
In order to calculate the transmission coefficient $|T|=|R_{\frac{1}{2}}|=|\sqrt{Q}{}_{\frac{1}{2}}{\mathcal R}|$ we need the value of the radial function at $r=2M$. The latter will be a superposition of ingoing and upgoing solutions (\ref{Rinupdefs}), i.e., 
\beq
{}_{\frac{1}{2}}{\mathcal R}={}_{\frac{1}{2}}{C}_{\rm (in)}{}_{\frac{1}{2}}{\mathcal R}_{\rm (in)}+{}_{\frac{1}{2}}{C}_{\rm (up)}{}_{\frac{1}{2}}{\mathcal R}_{\rm (up)}\,,
\eeq
where the coefficients ${}_{\frac{1}{2}}{C}_{\rm (in)}$ and ${}_{\frac{1}{2}}{C}_{\rm (up)}$ both depend on $l$ and $\omega$.
Their explicit form can be determined by using standard techniques (see, e.g., Refs. \cite{page76,mano}).
However, to first order in the acceleration parameter the correction turns out to be proportional to the Schwarzschild value
\beq
\Gamma_{\frac12}=\Gamma_{\frac12}^{\rm schw}\left(1-\frac{\eta}{2}E_1\right)
=M^2\omega^2\left(1-\frac25m\eta\right)\,,
\eeq
where $m=\pm1/2$.

Furthermore, one can evaluate the low-frequency (angle-averaged) absorption cross section 
\beq
\sigma_{\frac12}(\omega)=\frac{\pi}{\omega^2}\sum_{lm}\Gamma_{\frac12}
=2\pi M^2\,,\qquad M\omega\ll1\,,
\eeq
so that no corrections to the Schwarzschild result arise to that order.

\subsection{Two-component neutrinos}

Since neutrinos possess only one state of polarization, they can be described in terms of only two nonvanishing spinor components \cite{brill}.
Therefore, limiting our considerations to left-handed neutrinos (corresponding to the $P^A$ spinor), the Dirac current (\ref{gen-comp-curr}) becomes
\beq
J=\frac{\Omega^2}{r^2\sqrt{P}}[|f_1|^2 l + |f_2|^2 n + f_1f_2^{*} m + f_2f_1^{*} {\bar m}]\,.
\eeq
Substituting then the ansatz (\ref{sep-var}) leads to
\begin{equation}
\begin{array}{lcl}
J^t&=&\frac{\Omega^4}{2Q\sqrt{P}}\left(|R_{-\frac{1}{2}}|^2 |S_{-\frac{1}{2}}|^2 + |R_{\frac{1}{2}}|^2 |S_{\frac{1}{2}}|^2\right),\\[1ex]
J^r&=&\frac{\Omega^4}{2r^2\sqrt{P}}\left(|R_{-\frac{1}{2}}|^2 |S_{-\frac{1}{2}}|^2 - |R_{\frac{1}{2}}|^2 |S_{\frac{1}{2}}|^2\right),\\[1ex]
J^\theta&=&\frac{\Omega^4}{r^2\sqrt{Q}}\mbox{Re}\left(R_{-\frac{1}{2}}R_{\frac{1}{2}}^{*}\right) S_{-\frac{1}{2}} S_{\frac{1}{2}}^{*},\\[1ex]
J^\phi&=&-\frac{\Omega^4}{r^2\sqrt{Q}P\sin\theta}\mbox{Im}\left(R_{-\frac{1}{2}}R_{\frac{1}{2}}^{*}\right) S_{-\frac{1}{2}} S_{\frac{1}{2}}^{*}.
\end{array}
\end{equation}
Hence, in contrast to the case with $P^A$ and $\bar Q^{A'}$ both nonzero, if only one of the chiral components of the spinor field is considered, then $J$ is a null vector with nonvanishing angular components.

\section{Comparing acceleration and rotation effects}

We have solved in Section \ref{pert_ang} the eigenvalue problem associated with the angular equation given by Eqs. (\ref{calHdef})--(\ref{calHdef2}), to first order in the acceleration parameter $\eta$. The first order corrections to the energy eigenvalue are given by Eq. (\ref{E1sol_n}). 
It is interesting to compare the present analysis with the corresponding one for massless Dirac particles in a Kerr spacetime, in order to make a parallel between these two complementary situations of uniform background rotation and acceleration.

Let us briefly recall the Press and Teukolsky \cite{press} result in the Kerr case to first-order in the rotation parameter $a$.
The eigenvalue equation reads as ${\cal H}S=-ES$, where 
\beq\fl\qquad
\label{Hkerr}
{\cal H} \equiv \frac{1}{\sin\theta}\frac{\rmd }{\rmd \theta} 
  \left(\sin\theta \frac{\rmd}{\rmd \theta}\right) -\frac{m^2 +s^2 +2ms\cos\theta}{\sin^2\theta}
-2a\omega s\cos\theta,
\eeq
$S$ denotes the spheroidal harmonics and 
\beq
\label{Ekerr}
E=L-2a\omega \frac{s^2m}{L}
\eeq
is the energy eigenvalue.
Eq. (\ref{Ekerr}) shows that for a given value of $a\omega$ the energy depends quadratically on $s$, whereas for the C-metric it is linear in spin for fixed $\eta$ (see Eq. (\ref{E1sol_n})), so that in the latter case the energy levels turn out to split for particles with different spin.

We will show below that quite interestingly the two eigenvalue problems can be cast exactly in the same form, allowing to make the comparison easier. 
By introducing the new spin variable ${\tilde s}=s+m\eta$, the Hamiltonian operator (\ref{calHdef2}) can be written as 
\begin{eqnarray}\fl\qquad
\label{tildeHdef}
\tilde{\cal H} &\equiv &\frac{1}{\sin\theta}\frac{\rmd }{\rmd \theta} 
  \left(\sin\theta \frac{\rmd}{\rmd \theta}\right) 
-\frac{m^2 +{\tilde s}^2 +2m{\tilde s}\cos\theta}{\sin^2\theta}+\eta(m{\tilde s}+L\cos\theta),
\end{eqnarray}
so that the eigenvalue equation (\ref{calHdef2}) becomes $\tilde{\cal H}\,{}_{\tilde s}{\mathcal S}=-E\,{}_{\tilde s}{\mathcal S}$.
Notice that $L$ (or equivalently $l$) can in turn be suitably replaced by $L\to L+\eta L_1$ without changing the unperturbed Hamiltonian.
The constant term $m{\tilde s}$ in the first order operator (\ref{tildeHdef}) can then be reabsorbed by a redefinition of energy, i.e., $E\to E+\eta m{\tilde s}$.
This leads to a new eigenvalue problem $\bar{\cal H}\,{}_{\tilde s}\bar{\mathcal S}=-\bar E\,{}_{\tilde s}\bar{\mathcal S}$, with
\beq
\bar{\cal H}=\frac{1}{\sin\theta}\frac{\rmd }{\rmd \theta} 
  \left(\sin\theta \frac{\rmd}{\rmd \theta}\right) 
-\frac{m^2 +{\tilde s}^2 +2m{\tilde s}\cos\theta}{\sin^2\theta}+\eta L\cos\theta\,,
\eeq
which formally reproduces {\it exactly} the Kerr problem (\ref{Hkerr}) with the identifications $s\leftrightarrow{\tilde s}$ and $-2a\omega s\leftrightarrow\eta L$, giving in turn to the spheroidal wave equation a broader meaning.
The energy eigenvalue (\ref{E1sol_n}) can also be cast in the form (\ref{Ekerr}) by using the freedom in the choice of $L_1$.
As a result both angular eigenvalues and eigenfunctions in the C-metric case have a one-to-one correspondence with the Kerr energy spectrum and spin-weighted spheroidal harmonics.
A similar circumstance was observed in Ref. \cite{biniTN}, where the comparative analysis of the Teukolsky master equation in the Schwarzschild and Taub-NUT spacetimes revealed the (exact) correspondence of the spin-weight with a new spin variable depending on the NUT parameter $\ell$ according to $s\leftrightarrow s+2\omega\ell$.
The \lq\lq quantization'' property for the newly defined spin-weighted parameter $\tilde s$ (which must be a half-integer) thus implies a \lq\lq quantization'' property for the acceleration parameter $\eta$.
This genuinely (unexpected) new result for the acceleration has no counterpart with the rotation, for which no \lq\lq quantization'' rule applies in this context.

Finally, it is worth to mention that the couplings of the spin-weight with the rotation and with the acceleration (both of them inertial effects) have different origin. In fact, the spin-rotation coupling (also referred to as Mashhoon effect, see Ref. \cite{mash88}) appears as a {\it direct} coupling ($as$), whereas the acceleration always enters the Hamiltonian through the parameter $\eta=2MA$, so that any coupling with the spin is {\it indirect} ($AMs$), i.e., mediated by the mass of the black hole.
This is in agreement with the results of Ref. \cite{bcm04}.

\section{Concluding remarks}

We have investigated the behavior of massless Dirac particles in the spacetime of the vacuum C-metric, which describes the static region in the neighborhood a uniformly accelerating Schwarzschild black hole, under certain conditions.  
Because of the type D character of the solution, the Dirac equation is separable in a suitable coordinate system. 
We have adopted spherical-like coordinates and solved the one-dimensional radial and angular equations.   
We have considered first the angular part.
Treating the angular equation as an eigenvalue problem, we have computed the first-order corrections in a dimensionless parameter associated with the background acceleration with respect to the unperturbed Schwarzschild case. 
The solution for the angular eigenfunctions has been obtained as a series expansion in the basis of spin-weighted spherical harmonics.
The associated eigenvalues turn out to depend linearly in spin, differently from the case of a Kerr spacetime to first order in the rotation parameter, where the dependence is quadratic in spin. 
Therefore, the energy spectrum splits between particles with different spin due to the spacetime acceleration.
Concerning then the radial equation, we have explicitly computed the first terms of the post-Newtonian expansion to first order in the acceleration parameter.
We have analyzed the associated scattering problem and computed the corrections due to the acceleration to the particle absorption rate as well as to the angle-averaged cross section in the low-frequency limit with respect to the corresponding results for the Schwarzschild solution.
Furthermore, we have provided the exact solution for both radial and angular equations in terms of general Heun functions.
Finally, we have discussed the nature of the coupling between intrinsic spin and spacetime acceleration in comparison with the well known spin-rotation coupling in a Kerr spacetime, by writing the angular eigenvalue problem exactly in the same form as in the Kerr case (linearized in the rotation parameter).
This formal analogy implies a suitable redefinition of both the spin-weight $s$ and the quantum number $l$. We have thus identified a map relating the C-metric angular eigenfunctions and eigenvalues with the Kerr spheroidal harmonics and energy spectrum, implying a \lq\lq quantization'' rule for the acceleration parameter.

\appendix

\section{Perturbative angular solution: coefficients}
\label{CGcoeffs}

We list below the explicit expressions of the various quantities necessary to calculate the first-order corrections to both energy eigenvalues and eigenfunctions given by Eqs. (\ref{E1def}) and (\ref{Cdef3}).

The relevant Clebsch-Gordan coefficients to compute $\bra\cos\theta\ket_{l,l'}$ are
\begin{eqnarray}
\bra l 1\, m\, 0 | l-1 m\ket&=&-\sqrt{\frac{l^2-m^2}{l(2l+1)}}\,,\nonumber\\
\bra l 1\, m\, 0 | l m\ket&=&\frac{m}{\sqrt{l(l+1)}}\,,\nonumber\\
\bra l 1\, m\, 0 | l+1 m\ket&=&\sqrt{\frac{(l+1)^2-m^2}{(l+1)(2l+1)}}\,,
\end{eqnarray}
and
\begin{eqnarray}
\bra l 1\, -s 0 | l-1 -s\ket&=&-\frac12\sqrt{\frac{2l-1}{l}}\,,\nonumber\\
\bra l 1\, -s 0 | l -s\ket&=&-\frac{s}{\sqrt{l(l+1)}}\,,\nonumber\\
\bra l 1\, -s 0 | l+1 -s\ket&=&\frac12\sqrt{\frac{2l+3}{2l+1}}\,,
\end{eqnarray}
leading to 
\beq
\bra\cos\theta\ket_{l,l'}=\left\{
\begin{array}{cl}
\displaystyle\frac{\sqrt{l^2-m^2}}{2l}\,,&l'=l-1\\[3ex]
-\displaystyle\frac{ms}{l(l+1)}\,,&l'=l\\[3ex]
\displaystyle\frac{\sqrt{(l+1)^2-m^2}}{2(l+1)}\,,&l'=l+1\\
\end{array}
\right.
\,.
\eeq
Analogously, we find
\beq\fl\quad
\bra\cos^2\theta\ket_{l,l'}=\left\{
\begin{array}{cl}
\displaystyle\frac{\sqrt{l^2-m^2}\sqrt{(l-1)^2-m^2}}{4l(l-1)}\,,&l'=l-2\\[3ex]
-ms\displaystyle\frac{\sqrt{l^2-m^2}}{l(l-1)(l+1)}\,,&l'=l-1\\[3ex]
\displaystyle\frac{l(l+1)-m^2}{2l(l-1)(l+1)}\,,&l'=l\\[3ex]
-ms\displaystyle\frac{\sqrt{(l+1)^2-m^2}}{l(l+1)(l+2)}\,,&l'=l+1\\[3ex]
\displaystyle\frac{\sqrt{(l+1)^2-m^2}\sqrt{(l+2)^2-m^2}}{4(l+1)(l+2)}\,,&l'=l+2\\
\end{array}
\right.
\,,
\eeq
and
\beq\fl
\bra\cos^3\theta\ket_{l,l'}=\left\{
\begin{array}{cl}
\displaystyle\frac{\sqrt{l^2-m^2}\sqrt{(l-1)^2-m^2}\sqrt{(l-2)^2-m^2}}{8l(l-1)(l-2)}\,,&l'=l-3\\[3ex]
-3ms\displaystyle\frac{\sqrt{l^2-m^2}\sqrt{(l-1)^2-m^2}}{4l(l-2)(l-1)(l+1)}\,,&l'=l-2\\[3ex]
3\displaystyle\frac{\sqrt{l^2-m^2}(l^2-1-m^2)}{8l(l-1)(l+1)}\,,&l'=l-1\\[3ex]
-3ms\displaystyle\frac{[l(l+1)-1-m^2]}{2l(l-1)(l+1)(l+2)}\,,&l'=l\\[3ex]
3\displaystyle\frac{\sqrt{(l+1)^2-m^2}[l(l+2)-m^2]}{8l(l+1)(l+2)}\,,&l'=l+1\\[3ex]
-3ms\displaystyle\frac{\sqrt{(l+1)^2-m^2}\sqrt{(l+2)^2-m^2}}{4l(l+1)(l+2)(l+3)}\,,&l'=l+2\\[3ex]
\displaystyle\frac{\sqrt{(l+1)^2-m^2}\sqrt{(l+2)^2-m^2}\sqrt{(l+3)^2-m^2}}{8(l+1)(l+2)(l+3)}\,,&l'=l+3\\
\end{array}
\right.
\,.
\eeq

The first-order corrections to the energy eigenvalues (\ref{E1def}) then turn out to be
\beq
\label{E1sol}
E_1 = ms\frac{2(L^2+5m^2)-L(7+2m^2)}{(L-2)(L+m^2)}\,,
\eeq
whereas those to the coefficients $C^{l'}_{lm}$ (see Eq. (\ref{Cdef3})) are
\begin{eqnarray}\fl\qquad
C^{l-3}_{lm}&=&\frac{\sqrt{l^2-m^2}\sqrt{(l-1)^2-m^2}\sqrt{(l-2)^2-m^2}(l-3)(l+1)}{24(l-1)^2[(l+2)(l+3)+m^2]}\,,\nonumber\\
\fl\qquad
C^{l-2}_{lm}&=&\frac{\sqrt{l^2-m^2}\sqrt{(l-1)^2-m^2}[-2ms(l+1)+E_1(l-2)]}{4l(2l-1)[(l-2)(l-1)+m^2]}\,,\nonumber\\
\fl\qquad
C^{l-1}_{lm}&=&-\frac{\sqrt{l^2-m^2}\{l(l-1)[(l+1)^2-5m^2]-2m^2(l-5)+8msE_1\}}{8l(l+1)[l(l-1)+m^2]}\,,\nonumber\\
\fl\qquad
C^{l}_{lm}&=&0\,,\nonumber\\
\fl\qquad
C^{l+1}_{lm}&=&\frac{\sqrt{(l+1)^2-m^2}\{(l+1)(l+2)(l^2-5m^2)+2m^2(l+6)+8msE_1\}}{8l(l+1)[(l+1)(l+2)+m^2]}\,,\nonumber\\
\fl\qquad
C^{l+2}_{lm}&=&-\frac{l\sqrt{(l+1)^2-m^2}\sqrt{(l+2)^2-m^2}[(l+3)E_1-2lms]}{4(l+1)(2l+3)[(l+2)(l+3)+m^2]}\,,\nonumber\\
\fl\qquad
C^{l+3}_{lm}&=&-\frac{l(l+4)\sqrt{(l+1)^2-m^2}\sqrt{(l+2)^2-m^2}\sqrt{(l+3)^2-m^2}}{24(l+2)^2[(l+3)(l+4)+m^2]}\,.
\end{eqnarray}
Finally, the corrections to to the eigenfunctions then immediately follow from their definition (\ref{Edef}).

\section*{Acknowledgements}
All authors acknowledge ICRANet for partial support. DB and AG thank the INFN Sezione di Napoli for partial support. 
EB is financially supported by the CAPES-ICRANet program (BEX 13956/13-2).

\end{document}